\documentclass[prl, aps, english, twocolumn, superscriptaddress]{revtex4}

\usepackage{graphicx}

\usepackage{amssymb, amsmath, rotating}

\newcommand{\expec}[1]{\left < #1\right >}

\usepackage{amsmath}

\usepackage{graphicx}

\usepackage{rotating}

\newcommand{\lp}{\left ( }

\newcommand{\rp}{\right ) }

\newcommand{\beq}{\begin{eqnarray*}}

\newcommand{\eeq}{\end{eqnarray*}}

\newcommand{\be}{\begin{eqnarray}}

\newcommand{\ee}{\end{eqnarray}}

\newcommand{\mc}{\mathcal}

\def\lsim{\mathrel{\rlap{\lower4pt\hbox{\hskip1pt$\sim$}}

    \raise1pt\hbox{$<$}}}   

\def\gsim{\mathrel{\rlap{\lower4pt\hbox{\hskip1pt$\sim$}}

    \raise1pt\hbox{$>$}}}
    
\begin{document}

\title{Local versus global equilibration near the bosonic Mott-superfluid transition}

\author{Stefan S. Natu}

\email{ssn8@cornell.edu}

\affiliation{Laboratory of Atomic and Solid State Physics, Cornell University, Ithaca, New York 14853, USA.}

\author{Kaden R.~A. Hazzard}

\affiliation{JILA and University of Colorado, Boulder, Colorado 80309-0440, USA}

\author{Erich J. Mueller}

\affiliation{Laboratory of Atomic and Solid State Physics, Cornell University, Ithaca, New York 14853, USA.}

\begin{abstract}
We study the response of trapped two dimensional cold bosons to time dependent lattices. We find that in lattice ramps from $11$ (superfluid, $\hbar/U_{\text{i}} = 3$ms, $\hbar/J_{\text{i}} = 45$ms) to $16$ recoils (Mott, $\hbar/U_{\text{f}} = 2$ms, $\hbar/J_{\text{f}} = 130$ms)  the local number fluctuations remains at their equilibrium values if ramps are slower than $3$ ms. Global transport, however, is much slower ($1$s), especially in the presence of Mott shells.  This separation of timescales has practical implications for cold atom experiments and cooling protocols.
\end{abstract}

\maketitle

\textit{Introduction.---}Understanding and controlling the equilibration of cold atom systems is one of the most important current challenges in the field.  As isolated systems, the relaxation mechanisms are intrinsic and fundamental~\cite{kinoshita_quantum_2006,sadler_spontaneous_2006,hofferberth_non-equilibrium_2007,weiler_spontaneous_2008,du_observation_2008}.  The atomic systems are readily driven far from equilibrium, and are well suited for quantifying concepts of non-equilibrium dynamics~\cite{rigol_thermalization_2008, sengupta_quench_2004,kollath_quench_2007,zurek_dynamics_2005,polkovnikov_breakdown_2008,krutitsky_excitation_2010, oganesyan_energy_2009}.
 Moreover, controlling the equilibration of cold atoms is key to the next generation of experiments: for example one needs fast equilibration for condensed matter emulators~\cite{bloch_many-body_2008}.
Motivated by recent experiments~\cite{hung_slow_2010,bakr_probingsuperfluid-to-mott-insulator_2010,sherson_single-atom_2010}, we conduct numerical simulations of the response of a gas of bosons to a change in the intensity of an applied optical lattice.

 Despite being performed under similar conditions, three recent experiments~\cite{hung_slow_2010,bakr_probingsuperfluid-to-mott-insulator_2010,sherson_single-atom_2010} find relaxation rates for two-dimensional lattice bosons that differ by two orders of magnitude. Here we show that these discrepancies can be explained by a separation of timescales for local equilibration and global transport.  We illustrate this result by numerical simulations within a time-dependent Gutzwiller mean-field theory.  We further explore the parameters, such as system size and trap geometry, which influence these timescales.

The separation of timescales for local and global equilibrium is unsurprising, and emerges in most interacting systems and materials.  For example, in the air around us, local equilibrium is achieved on the collision time ($\sim$ns), but global equilibrium is limited by transport coefficients and is relatively slow.  Typically one expects the slow variables to be those that are conserved (such as density and energy density) and those which correspond to broken symmetries (such as the phase of the superfluid order parameter).  Although we do not do so here, integrating out the fast degrees of freedom leaves ``hydrodynamic" equations for the slow degrees of freedom.  The form of these hydrodynamic equations are strongly constrained by symmetries, allowing phenomenological descriptions~\cite{landau_fluid_1987,pitaevskii_physical_1981}.

A practical consequence of this separation of timescales is that adiabaticity is much easier to maintain if one changes parameters in such a way that very little mass transport is necessary -- a principle which is widely used in cold atom experiments.

\begin{figure}

\begin{picture}(200, 105)
\put(0, -10){\includegraphics[scale=0.62]{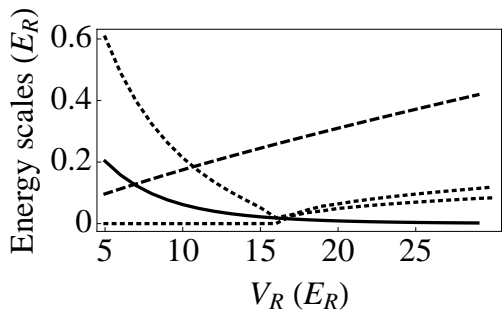}}
\end{picture}

\caption{\label{fig:params}\textbf{Energy scales as a function of lattice depth}: Microscopic parameters in the $2D$ Bose-Hubbard Hamiltonian (Eq.\ref{eq:1}):~$4J$ (solid), and $U$ (dashed) as a function of lattice depth \cite{jaksch_cold_1998} for $^{87}$Rb in a $d = 680$ nm lattice. The dotted curves are the two lowest $k=0$ excitations from linearizing Eq.~(\ref{eq:2}) at unity filling. In the superfluid state, the Goldstone mode has zero energy. In the Mott state, these are the particle/hole excitations.}

\end{figure}

\textit{Theoretical Setup.---}Bosonic atoms trapped by interfering laser beams are well described by the Bose Hubbard Hamiltonian~\cite{fisher_boson_1989}
\begin{equation}\label{eq:1}
{\cal{H}} = -J\sum_{\langle ij \rangle}(a^{\dagger}_{i}a_{j} + h.c) +\sum_{i}\left( \frac{U}{2}n_{i}(n_{i}-1)
- \mu^{\text{i}}n_{i}\right)
\end{equation}
where $a$ and $a^\dagger$ are bosonic annihilation and creation operators, $J$ is the tunneling, and $U$ is on-site interaction. We denote $\mu^{i} = \mu - V_{\text{ex}}(i)$, where $\mu$ is the chemical potential and $V_{\text{ex}}(i)$ is the external potential at site $i$~\cite{jaksch_cold_1998}.  The first sum is over all nearest neighbor sites in the plane.  In Figure~\ref{fig:params}, we show $U$ and $J$ as a function of lattice depth $V_R$ for  $^{87}$Rb in a $d=680$nm lattice generated by light of wavelength $\lambda=1360$nm. For deep lattices, $U= \sqrt{8/\pi} (ka_{s})E_R \sqrt{V_R/E_R}(V_{Rz}/E_R)^{1/4}$ and $J = (4/\sqrt{\pi}) (E_R V_R^{3})^{1/4}\exp \lp -2 \sqrt{V_R/E_R}\rp$, where $E_R = k^{2}/2m$ is the lattice recoil energy in terms of the light wave-vector $k = 2\pi/\lambda$ for light with wavelength $\lambda$, $V_R, V_{Rz}$ are the radial and axial lattice depths, $a_s$ is the scattering length~\cite{zwerger_mott-hubbard_2003}.  Different two-dimensional experiments use different strengths of axial confinement ($V_{Rz}$). Since $U$ only depends on $V^{1/4}_{Rz}$, we will make the simplest choice, $V_{Rz} = V_{R}$. None of our conclusions are qualitatively affected by this assumption.

We calculate dynamics using a time dependent Gutzwiller ansatz~\cite{fisher_boson_1989}, which approximates the wavefunction by $\Psi = \bigotimes_{i}\sum_{m}c_{m}^{(i)}(t)|m\rangle_i$ where $|m\rangle_i$ is the $m$-particle Fock state on site $i$, and the coefficients $c_{m}^{(i)}(t)$ are generally space and time dependent.

This mean-field ansatz reduces Eq.~\ref{eq:1} to a sum of single site Hamiltonians ${\cal{H}}_{i} = -4t(\langle \alpha_i\rangle^* a_i + \langle \alpha_i\rangle a_i^{\dagger}) +  4 t |\langle \alpha_i \rangle|^2 + \frac{U}{2}n_{i}(n_{i}-1) - [\mu-V(i)] n_{i}$ at each site $i$. Truncating the basis at each site to a maximum $M$ particles, ${\cal{H}}_{i}$ is an $(M+1)\times (M+1)$ matrix at each site, and depends on the other sites only through $\langle \alpha_{i} \rangle = (1/4)\sum_{\langle j \rangle}\langle a_{j} \rangle$, where $\expec{a_{j}}=\sum_m \sqrt{m+1}c_{m+1}^{(j)}c_m^{(j)}$, and the sum over $j$ includes all four nearest neighbor sites.

Schr\"odinger's equation $i\partial_{t}\psi = {\cal{H}}\psi$ for $\Psi$ yields a set of differential equations for  the $c^{i}_{m}$:
\begin{eqnarray}\label{eq:2}
i\partial_{t}c^{i}_{m}(t) = -4J(t)(\langle \alpha_{i}\rangle^{*}\sqrt{m+1}c^{i}_{m+1}  +  \langle \alpha_{i} \rangle \sqrt{m}c^{i}_{m-1}) +\hspace{-6mm}\nonumber\\ \left(\frac{U(t)}{2}m(m-1) - \mu^{i} m +4 J(t)|\langle \alpha_{i} \rangle|^2 \right)c^{i}_{m}
\end{eqnarray}
The tunnelings and on-site interactions are dynamically tuned by changing the lattice depth in time ($t$). We study population dynamics across the superfluid-insulator transition by ramping the lattice linearly in time using the protocol $V(t) = V_{\text{i}} +   (V_{\text{f}}-V_{\text{i}})(t/\tau_\text{r})$, where $V_{\text{i}}$ and $V_{\text{f}}$ are the initial and final lattice depths, and $\tau_\text{r}$ is the ramp time. We consider a time independent  radially symmetric harmonic trap, $V_{\text{ex}}  = \frac{1}{2}m\omega^{2}(x^{2} + y^{2})$. 

We approximate the ground state by finding the stationary solution to Eq.~(\ref{eq:2}), $c_m^{i}(t)=e^{-i\epsilon t} c_m^{i}$, where $\epsilon$ can be identified with the energy per site.  We use an iterative algorithm, starting with a trial $\alpha_i$, then find $c_m^{i}$ by solving the eigenvalue problem in Eq.~(\ref{eq:2}).  We calculate a new $\alpha_i$ and repeat until the subsequent change in $\alpha_i$ is sufficiently small.  To calculate time dynamics,
 we use a split-step method~\cite{press_numerical_2007} and sequential site updates~\cite{wernsdorfer_lattice-ramp-induced_2010}.  This approach conserves both total particle number and energy (for time-independent Hamiltonians).
 
The resulting dynamics describe the behavior of a single quantum state, rather than a density matrix.  Nevertheless, the equations governing the time dependent Gutzwiller ansatz are highly nonlinear and contain a large number of degrees of freedom.  This structure is rich enough that under appropriate conditions time dynamics leads to thermalization, with (on average) energy equally distributed among all modes. 

\textit{Results.---}We consider several different scenarios in order to fully explore the response this system to a lattice ramp.  We start by analyzing a homogeneous system: this investigation yields the timescale for local equilibration.  This timescale sets the fundamental limit for how fast equilibration can take place in the absence of global mass transport. Similar to the Harvard experiments~\cite{bakr_probingsuperfluid-to-mott-insulator_2010}, we find that local equilibrium can be maintained even under relatively rapid quenches through the superfluid-Mott boundary.  

Next we explore the requirements for maintaining global equilibrium. We show that equilibration times are much longer in systems requiring large amounts of particle transport.   This situation is exacerbated by the presence of large Mott domains, as in the Chicago experiments \cite{hung_slow_2010}.

We conclude by showing that even in large systems, rapid \textit{global} equilibration can be achieved, if the trap parameters are chosen in a way as to minimize transport between intervening Mott shells. Our results in this section are consistent with the Munich experiments \cite{sherson_single-atom_2010}. 

\textit{Local equilibration.---}In an isolated homogeneous system, ramping the depth of an optical lattice does not lead to bulk mass transport.  Instead, all of the temporal dynamics simply involve the evolution of number fluctuations and correlations.  Thus 
equilibration is governed by local physics and Eq.~\ref{eq:2} reduces to the single site problem.  We numerically integrate this nonlinear set of ordinary differential equations, taking $J$ and $U$ functions of time, corresponding to a linear ramp of the lattice from depth $V_\text{i}$ to $V_\text{f}$.  We vary $V_\text{i}$, $V_\text{f}$, and the ramp time $\tau_{\text{r}}$. We take all parameters to correspond to $^{87}$Rb atoms, and take $n=1$ particles per site.

At unity filling, near the Mott transition, we truncate the basis to at most 2 particles per site. In this truncated basis, the probability of having a single particle per site $P(1)$ is identical to the probability of having an odd number of particles per site, which is the experimental observable in the Harvard experiments \cite{bakr_probingsuperfluid-to-mott-insulator_2010}.

Both of the gapped $q=0$ single-particle excitations (see Fig.\ref{fig:params} and Ref.\cite{menotti_spectral_2008-2}), and the continuum of two-phonon excitations contribute to the non-adiabatic evolution. All of these  modes  are captured in a time-dependent Gutzwiller framework \cite{krutitsky_excitation_2010}.  One expects that the number of excitations goes to zero as the ramp rate vanishes. When gapped excitations of energy $\Delta$ dominate the dissipation, then the ramp becomes adiabatic when $\frac{1}{\Delta^{2}}d\Delta/dt \ll 1$ \cite{landau_quantum_1981}.

\begin{figure}
\begin{picture}(100, 100)
\put(-50, -5){\includegraphics[scale=0.55]{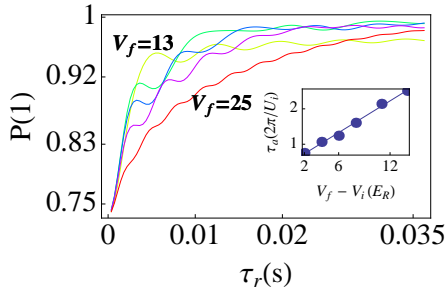}}
\end{picture}

\caption{\label{fig:homogresults}(Color Online) \textbf{Population dynamics at unity density $n=1$} (Top): Probability of having one particle per site at the end of a lattice ramp from $V_\text{i}=11E_R$ lattice to (top to bottom) $V_\text{f}=13$(yellow), $15$(green), $17$(blue), $19$(purple) and $25$(red) in units of  $E_{R}$ after different lattice ramp times $\tau_\text{r}= 0.1/U_{\text{i}} \sim 0.3$ms to $10/U_{\text{i}}$. Inset: Fitting these curves to simple exponentials yields a fast timescale for lattice equilibration of $\tau_{\text{a}} \sim \frac{2\pi}{U_{\text{i}}}$. The best fit line is shown as a guide to the eye.} 
\end{figure}

In Fig.~\ref{fig:homogresults} we show that the timescale for local equilibration is very short. Starting with a superfluid at $V_\text{i}=11E_R$, we ramp up to different lattice depths. We plot the time evolution of the probability that a single particle sits at a given site as we vary the 
the ramp time $\tau_\text{r}$  from $0.1 \hbar/U_{\text{i}}$ to $10  \hbar/U_{\text{i}}$, where $U_{\text{i}} = \hbar/3$ms. Our scheme is identical to that considered by the Harvard experiments \cite{bakr_probingsuperfluid-to-mott-insulator_2010}. Fitting these curves to simple exponentials yields a characteristic timescale $\tau_\text{a}$, which as we show in the inset, is comparable to $U^{-1}_{\text{i}}$.

\textit{Inhomogeneous dynamics.---}We now consider an inhomogenous system by imposing a harmonic external potential on top of the lattice. The protocol for lattice ramps is same as before, starting with a superfluid at $11$ recoil lattice depth. The central chemical potential is chosen such that the central density is close to unity, justifying the truncated basis ($M=2$) used here. Throughout we define time in units of $2\pi/U_{\text{i}}$ where $U_{\text{i}}$ is the on-site interaction at $V_\text{i} =11E_R$ equal to $\sim 2\pi \times 300$Hz. We use a trapping frequency $\omega = 15$Hz.

\begin{figure}[tbp]

\begin{picture}(150, 120)

\put(-50, 0){\includegraphics[scale=0.37]{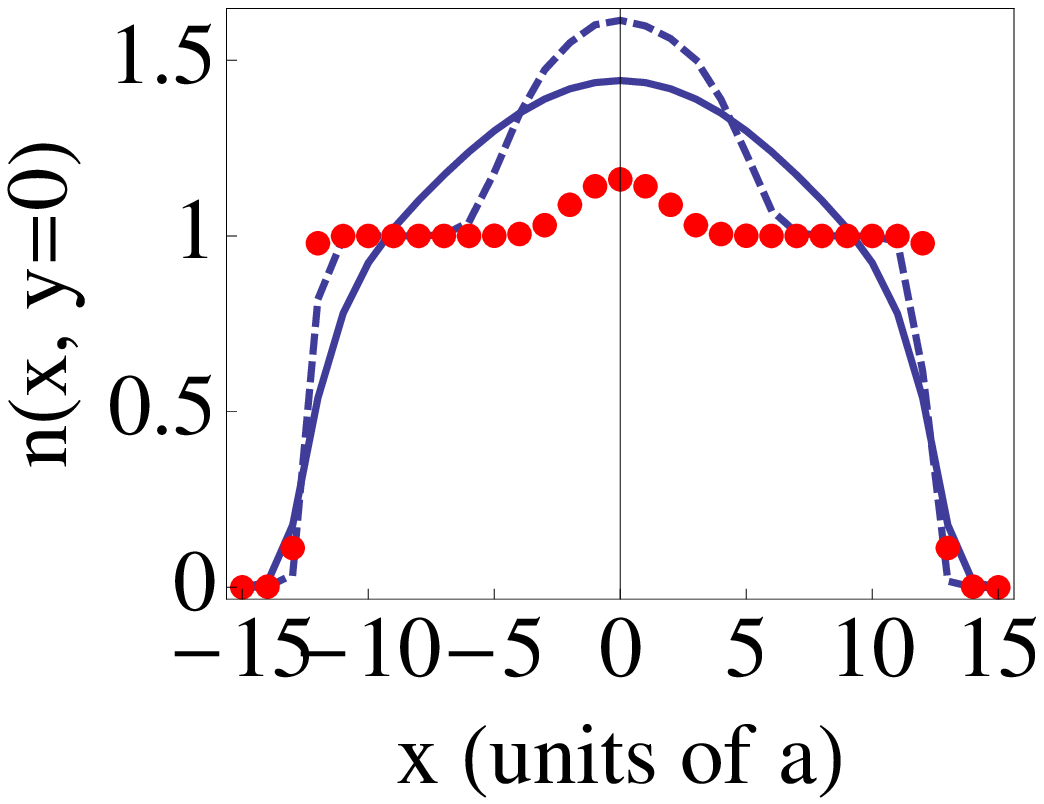}}

\put(70, 0){\includegraphics[scale=0.35]{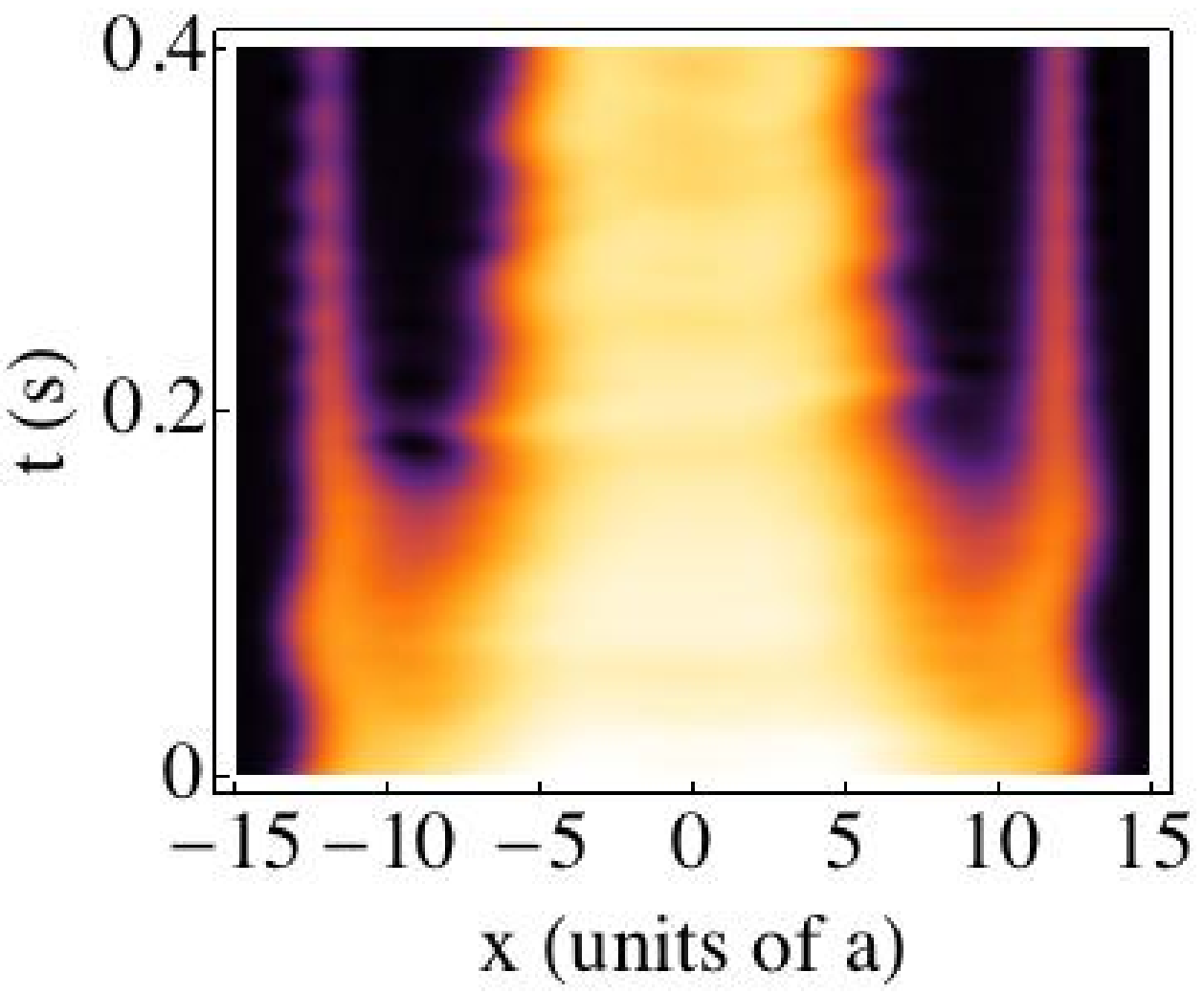}}

\end{picture}

\caption{\label{fig:-1}(Color Online) ~\textbf{Slow transport across Mott region} (Left) Evolution of an initial superfluid state (solid) at $V_{\text{i}} = 11E_R$ and $N = 500$ in a $15$Hz radial trapping potential. Final density profile (dashed) after a ramp $\tau_{r} = 120 \times 2\pi/U_{\text{i}} \sim 400$ ms, is very different from the equilibrium state (dotted) at $V_{\text{f}}  = 16E_R$. (Right) Density plot showing the time evolution of the coherences ($\mc C_{i}\equiv -\langle a_{i} \rangle \sum_{j}\langle a^{*}_{j} \rangle$), a growing Mott region in the wings which cuts off transport in the intervening superfluid producing a non-equilibrium final state at late times. Brighter colors correspond to higher coherence}

\end{figure}

In Fig.~\ref{fig:-1} we plot the density profile after a lattice ramp from $V_{\text{i}} = 11E_R$ to $V_{\text{f}}= 16E_R$ in a time $t = 120 \times 2\pi/U_{\text{i}}$ for a system $30\times 30$ sites containing $500$ particles. As shown previously, this ramp is sufficiently slow to be locally adiabatic. The parameters are chosen such that at later times, a large Mott region separates the central superfluid from the edge in the final state. In the Chicago experiment, \cite{hung_slow_2010}, this Mott domain was $\sim 50$ sites wide while in our simulations it is $\sim 8$ sites.

Like the experiment, we find that after this time the density profile of the final state (dashed line) is very different from the equilibrium state at $V_{\text{f}}$(dotted red), implying a relaxation time much longer the ramp time of $400$ms. Indeed, further simulations show that it is longer than the experimental timescale of seconds. In the remainder of this section we describe the cause of the slow equilibration, and conduct a number of additional simulations to illustrate how equilbration times depend on the various experimental parameters.

The major bottleneck for equilibration in Fig.~\ref{fig:-1} is mass transport across the Mott region.  To illustrate the spatial location of the Mott insulator, in  Fig.~\ref{fig:-1}(b)  we plot the coherences $\mc C_{i}\equiv -\langle a_{i} \rangle \sum_{j}\langle a^{*}_{j} \rangle$ as a function of time, where $i, j$ denote nearest neighbor pairs. Mott regions ($\mc C=0$) show up as dark regions in the density plot. The Mott plateau widens over time, isolating the central superfluid. The peak atomic density in the initial lattice exceeds that of the equilibrium state at the final lattice depth. However the Mott region prevents mass flow from the center to the edge. The exponential suppression of transport through the Mott region was studied by Vishveshwara and Lannert \cite{vishveshwara_josephson_2008}. 

\begin{figure}[tbp]

\begin{picture}(150, 120)

\put(-50, 0){\includegraphics[scale=0.37]{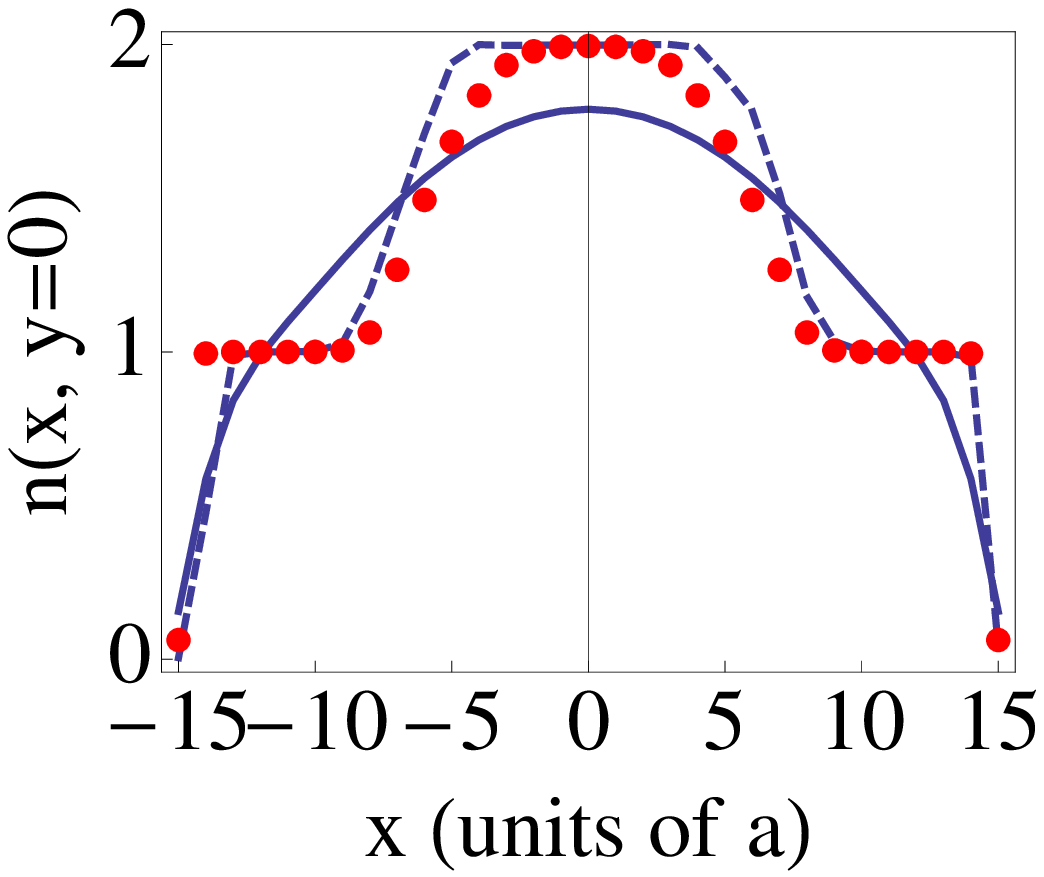}}

\put(70, 0){\includegraphics[scale=0.35]{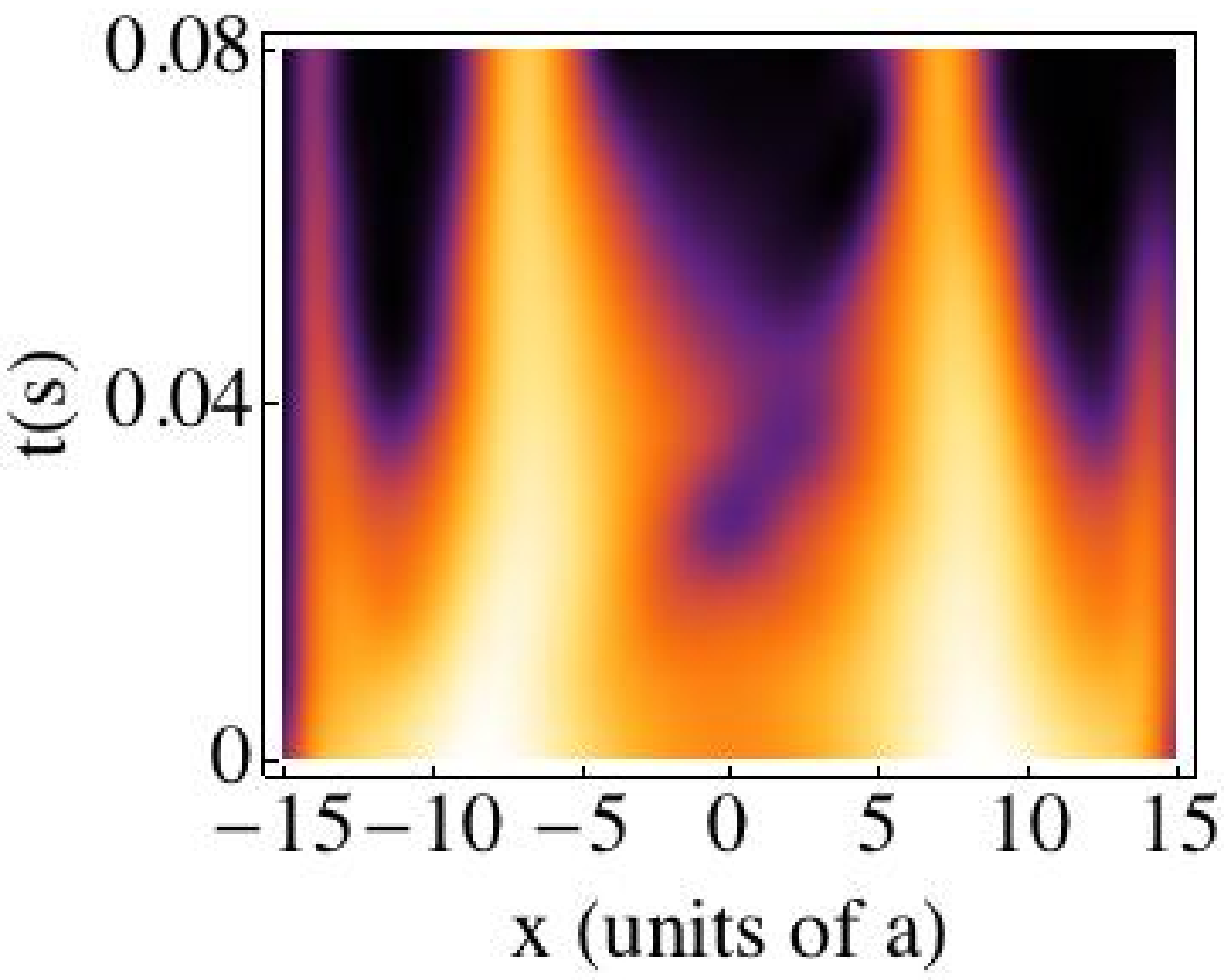}}

\end{picture}

\caption{\label{fig:-3}(Color Online)~\textbf{Time-evolution at higher density}(Left): Evolution of an initial superfluid state for $V_{\text{i}} = 11E_R$ and $N = 900$ (solid) in a $15$Hz radial trapping potential in a linear ramp with $\tau_\text{r} = 25 \times 2\pi/U_{\text{i}} = 80$ms. The dotted profile is the $T=0$ equilibrium Gutzwiller profile at $V_{0}  = 16E_R$ for the same parameters. The final density profile (dashed) agrees with the $T=0$ equilibrium Gutzwiller profile. (Right) Time evolution of the spatial coherence distribution, showing the formation of an $n=1$ and $n=2$ Mott plateaus. Lighter colors imply larger coherences.}

\end{figure}

\textit{Fast equilibration without transport.---}Here we show that equilibration times can be dramatically reduced when parameters are chosen such that no bulk transport across Mott regions is required. The parameters are chosen to mimic the large systems considered by Sherson \textit{et al.} \cite{sherson_single-atom_2010}, which attained \textit{global} equilibrium on timescales comparable to $100$ms. Figure~\ref{fig:-3} shows the time-evolution of an initial state at $V_{\text{i}} =11$ at $N = 800$ in  $2.5$Hz trap, and a central chemical potential of $1.4U$. We find that after an evolution of $\tau_{\text{r}} = 25 \times 2\pi/U_{\text{i}}$, the final profile (dashed) is close to the equilibrium $T=0$ Gutzwiller prediction (dotted).

Despite the fact that the $n=1$ Mott region is of similar size, we find faster equilibration times in this system as compared to the one in Fig.\ref{fig:-1}. The difference is that here parameters are chosen such that the total number of particles in the center is the same in the initial and final states. Thus no transport is needed across the Mott reion.

\textit{Summary.---}Our work was motivated by three experiments. The Chicago experiments \cite{hung_slow_2010} have a large system, and an extremely wide Mott region ($\sim 50$ sites) inhibiting transport. We showed by smaller scale calculations that once the Mott shell is a significant fraction of the system size, the dynamics slow dramatically, leading to extremely long equilibration times observed in the experiment. 

The Harvard experiments \cite{bakr_probingsuperfluid-to-mott-insulator_2010} have good evidence of  local equilibrium on short timescales. We investigate this timescale by studying a homogenous system. One insight into the timescale for equilibration is that the lowest energy $k=0$ single particle excitation has a gap $\sim U$. This energy scale appears to set the timescale for adiabaticity. 

The Munich experiments \cite{sherson_single-atom_2010} find excellent agreement with equilibrium profiles after very short  $75$ms ($\sim J$) ramps. For the parameters we consider, our simulations reproduce this result. We attribute the difference between the Chicago and Munich observations to the greater amount of transport accross Mott regions required to reach equilibrium for the Chicago parameters. We believe that although the Harvard experiments have good evidence for local equilibration on timescales of $1/U$, they may be out of global equilibrium after such short times.

\textit{Acknowledgements.---}This work was supported in part by a grant from the Army Research office with funding from the DARPA OLE Program. SN thanks S. K. Baur for insightful discussions. KRAH thanks Ana Maria Rey for useful discussions and support.


\begin{thebibliography}{26}
\expandafter\ifx\csname natexlab\endcsname\relax\def\natexlab#1{#1}\fi
\expandafter\ifx\csname bibnamefont\endcsname\relax
  \def\bibnamefont#1{#1}\fi
\expandafter\ifx\csname bibfnamefont\endcsname\relax
  \def\bibfnamefont#1{#1}\fi
\expandafter\ifx\csname citenamefont\endcsname\relax
  \def\citenamefont#1{#1}\fi
\expandafter\ifx\csname url\endcsname\relax
  \def\url#1{\texttt{#1}}\fi
\expandafter\ifx\csname urlprefix\endcsname\relax\def\urlprefix{URL }\fi
\providecommand{\bibinfo}[2]{#2}
\providecommand{\eprint}[2][]{\url{#2}}

\bibitem[{\citenamefont{Kinoshita et~al.}(2006)\citenamefont{Kinoshita, Wenger,
  and Weiss}}]{kinoshita_quantum_2006}
\bibinfo{author}{\bibfnamefont{T.}~\bibnamefont{Kinoshita}},
  \bibinfo{author}{\bibfnamefont{T.}~\bibnamefont{Wenger}}, \bibnamefont{and}
  \bibinfo{author}{\bibfnamefont{D.~S.} \bibnamefont{Weiss}},
  \bibinfo{journal}{Nature} \textbf{\bibinfo{volume}{440}},
  \bibinfo{pages}{900} (\bibinfo{year}{2006}), ISSN \bibinfo{issn}{0028-0836}.

\bibitem[{\citenamefont{Sadler et~al.}(2006)\citenamefont{Sadler, Higbie,
  Leslie, Vengalattore, and {Stamper-Kurn}}}]{sadler_spontaneous_2006}
\bibinfo{author}{\bibfnamefont{L.~E.} \bibnamefont{Sadler}},
  \bibinfo{author}{\bibfnamefont{J.~M.} \bibnamefont{Higbie}},
  \bibinfo{author}{\bibfnamefont{S.~R.} \bibnamefont{Leslie}},
  \bibinfo{author}{\bibfnamefont{M.}~\bibnamefont{Vengalattore}},
  \bibnamefont{and} \bibinfo{author}{\bibfnamefont{D.~M.}
  \bibnamefont{{Stamper-Kurn}}}, \bibinfo{journal}{Nature}
  \textbf{\bibinfo{volume}{443}}, \bibinfo{pages}{312} (\bibinfo{year}{2006}),
  ISSN \bibinfo{issn}{0028-0836}.

\bibitem[{\citenamefont{Hofferberth et~al.}(2007)\citenamefont{Hofferberth,
  Lesanovsky, Fischer, Schumm, and
  Schmiedmayer}}]{hofferberth_non-equilibrium_2007}
\bibinfo{author}{\bibfnamefont{S.}~\bibnamefont{Hofferberth}},
  \bibinfo{author}{\bibfnamefont{I.}~\bibnamefont{Lesanovsky}},
  \bibinfo{author}{\bibfnamefont{B.}~\bibnamefont{Fischer}},
  \bibinfo{author}{\bibfnamefont{T.}~\bibnamefont{Schumm}}, \bibnamefont{and}
  \bibinfo{author}{\bibfnamefont{J.}~\bibnamefont{Schmiedmayer}},
  \bibinfo{journal}{Nature} \textbf{\bibinfo{volume}{449}},
  \bibinfo{pages}{324} (\bibinfo{year}{2007}), ISSN \bibinfo{issn}{0028-0836}.

\bibitem[{\citenamefont{Weiler et~al.}(2008)\citenamefont{Weiler, Neely,
  Scherer, Bradley, Davis, and Anderson}}]{weiler_spontaneous_2008}
\bibinfo{author}{\bibfnamefont{C.~N.} \bibnamefont{Weiler}},
  \bibinfo{author}{\bibfnamefont{T.~W.} \bibnamefont{Neely}},
  \bibinfo{author}{\bibfnamefont{D.~R.} \bibnamefont{Scherer}},
  \bibinfo{author}{\bibfnamefont{A.~S.} \bibnamefont{Bradley}},
  \bibinfo{author}{\bibfnamefont{M.~J.} \bibnamefont{Davis}}, \bibnamefont{and}
  \bibinfo{author}{\bibfnamefont{B.~P.} \bibnamefont{Anderson}},
  \bibinfo{journal}{Nature} \textbf{\bibinfo{volume}{455}},
  \bibinfo{pages}{948} (\bibinfo{year}{2008}), ISSN \bibinfo{issn}{0028-0836}.

\bibitem[{\citenamefont{Du et~al.}(2008)\citenamefont{Du, Luo, Clancy, and
  Thomas}}]{du_observation_2008}
\bibinfo{author}{\bibfnamefont{X.}~\bibnamefont{Du}},
  \bibinfo{author}{\bibfnamefont{L.}~\bibnamefont{Luo}},
  \bibinfo{author}{\bibfnamefont{B.}~\bibnamefont{Clancy}}, \bibnamefont{and}
  \bibinfo{author}{\bibfnamefont{J.~E.} \bibnamefont{Thomas}},
  \bibinfo{journal}{Phys. Rev. Lett.} \textbf{\bibinfo{volume}{101}},
  \bibinfo{pages}{150401} (\bibinfo{year}{2008}).

\bibitem[{\citenamefont{Rigol et~al.}(2008)\citenamefont{Rigol, Dunjko, and
  Olshanii}}]{rigol_thermalization_2008}
\bibinfo{author}{\bibfnamefont{M.}~\bibnamefont{Rigol}},
  \bibinfo{author}{\bibfnamefont{V.}~\bibnamefont{Dunjko}}, \bibnamefont{and}
  \bibinfo{author}{\bibfnamefont{M.}~\bibnamefont{Olshanii}},
  \bibinfo{journal}{Nature} \textbf{\bibinfo{volume}{452}},
  \bibinfo{pages}{854} (\bibinfo{year}{2008}), ISSN \bibinfo{issn}{0028-0836}.

\bibitem[{\citenamefont{Sengupta et~al.}(2004)\citenamefont{Sengupta, Powell,
  and Sachdev}}]{sengupta_quench_2004}
\bibinfo{author}{\bibfnamefont{K.}~\bibnamefont{Sengupta}},
  \bibinfo{author}{\bibfnamefont{S.}~\bibnamefont{Powell}}, \bibnamefont{and}
  \bibinfo{author}{\bibfnamefont{S.}~\bibnamefont{Sachdev}},
  \bibinfo{journal}{Phys. Rev. A} \textbf{\bibinfo{volume}{69}},
  \bibinfo{pages}{053616} (\bibinfo{year}{2004}).

\bibitem[{\citenamefont{Kollath et~al.}(2007)\citenamefont{Kollath,
  L{\"a}uchli, and Altman}}]{kollath_quench_2007}
\bibinfo{author}{\bibfnamefont{C.}~\bibnamefont{Kollath}},
  \bibinfo{author}{\bibfnamefont{A.~M.} \bibnamefont{L{\"a}uchli}},
  \bibnamefont{and} \bibinfo{author}{\bibfnamefont{E.}~\bibnamefont{Altman}},
  \bibinfo{journal}{Phys. Rev. Lett.} \textbf{\bibinfo{volume}{98}},
  \bibinfo{pages}{180601} (\bibinfo{year}{2007}).

\bibitem[{\citenamefont{Zurek et~al.}(2005)\citenamefont{Zurek, Dorner, and
  Zoller}}]{zurek_dynamics_2005}
\bibinfo{author}{\bibfnamefont{W.~H.} \bibnamefont{Zurek}},
  \bibinfo{author}{\bibfnamefont{U.}~\bibnamefont{Dorner}}, \bibnamefont{and}
  \bibinfo{author}{\bibfnamefont{P.}~\bibnamefont{Zoller}},
  \bibinfo{journal}{Phys. Rev. Lett.} \textbf{\bibinfo{volume}{95}},
  \bibinfo{pages}{105701} (\bibinfo{year}{2005}).

\bibitem[{\citenamefont{Polkovnikov and
  Gritsev}(2008)}]{polkovnikov_breakdown_2008}
\bibinfo{author}{\bibfnamefont{A.}~\bibnamefont{Polkovnikov}} \bibnamefont{and}
  \bibinfo{author}{\bibfnamefont{V.}~\bibnamefont{Gritsev}},
  \bibinfo{journal}{Nat Phys} \textbf{\bibinfo{volume}{4}},
  \bibinfo{pages}{477} (\bibinfo{year}{2008}), ISSN \bibinfo{issn}{1745-2473}.

\bibitem[{\citenamefont{Krutitsky and Navez}(2010)}]{krutitsky_excitation_2010}
\bibinfo{author}{\bibfnamefont{K.~V.} \bibnamefont{Krutitsky}}
  \bibnamefont{and} \bibinfo{author}{\bibfnamefont{P.}~\bibnamefont{Navez}},
  \bibinfo{journal}{arxiv:1004.2121}  (\bibinfo{year}{2010}),
  \urlprefix\url{http://arxiv.org/abs/1004.2121}.

\bibitem[{\citenamefont{Oganesyan et~al.}(2009)\citenamefont{Oganesyan, Pal,
  and Huse}}]{oganesyan_energy_2009}
\bibinfo{author}{\bibfnamefont{V.}~\bibnamefont{Oganesyan}},
  \bibinfo{author}{\bibfnamefont{A.}~\bibnamefont{Pal}}, \bibnamefont{and}
  \bibinfo{author}{\bibfnamefont{D.~A.} \bibnamefont{Huse}},
  \bibinfo{journal}{Phys. Rev. B} \textbf{\bibinfo{volume}{80}},
  \bibinfo{pages}{115104} (\bibinfo{year}{2009}).

\bibitem[{\citenamefont{Bloch et~al.}(2008)\citenamefont{Bloch, Dalibard, and
  Zwerger}}]{bloch_many-body_2008}
\bibinfo{author}{\bibfnamefont{I.}~\bibnamefont{Bloch}},
  \bibinfo{author}{\bibfnamefont{J.}~\bibnamefont{Dalibard}}, \bibnamefont{and}
  \bibinfo{author}{\bibfnamefont{W.}~\bibnamefont{Zwerger}},
  \bibinfo{journal}{Rev. Mod. Phys.} \textbf{\bibinfo{volume}{80}},
  \bibinfo{pages}{885} (\bibinfo{year}{2008}).

\bibitem[{\citenamefont{Hung et~al.}(2010)\citenamefont{Hung, Zhang, Gemelke,
  and Chin}}]{hung_slow_2010}
\bibinfo{author}{\bibfnamefont{C.}~\bibnamefont{Hung}},
  \bibinfo{author}{\bibfnamefont{X.}~\bibnamefont{Zhang}},
  \bibinfo{author}{\bibfnamefont{N.}~\bibnamefont{Gemelke}}, \bibnamefont{and}
  \bibinfo{author}{\bibfnamefont{C.}~\bibnamefont{Chin}},
  \bibinfo{journal}{Phys. Rev. Lett.} \textbf{\bibinfo{volume}{104}},
  \bibinfo{pages}{160403} (\bibinfo{year}{2010}).

\bibitem[{\citenamefont{Bakr et~al.}(2010)\citenamefont{Bakr, Peng, Tai, Ma,
  Simon, Gillen, Folling, Pollet, and
  Greiner}}]{bakr_probingsuperfluid-to-mott-insulator_2010}
\bibinfo{author}{\bibfnamefont{W.~S.} \bibnamefont{Bakr}},
  \bibinfo{author}{\bibfnamefont{A.}~\bibnamefont{Peng}},
  \bibinfo{author}{\bibfnamefont{M.~E.} \bibnamefont{Tai}},
  \bibinfo{author}{\bibfnamefont{R.}~\bibnamefont{Ma}},
  \bibinfo{author}{\bibfnamefont{J.}~\bibnamefont{Simon}},
  \bibinfo{author}{\bibfnamefont{J.~I.} \bibnamefont{Gillen}},
  \bibinfo{author}{\bibfnamefont{S.}~\bibnamefont{Folling}},
  \bibinfo{author}{\bibfnamefont{L.}~\bibnamefont{Pollet}}, \bibnamefont{and}
  \bibinfo{author}{\bibfnamefont{M.}~\bibnamefont{Greiner}},
  \bibinfo{journal}{Science} p. \bibinfo{pages}{science.1192368}
  (\bibinfo{year}{2010}).

\bibitem[{\citenamefont{Sherson et~al.}(2010)\citenamefont{Sherson, Weitenberg,
  Endres, Cheneau, Bloch, and Kuhr}}]{sherson_single-atom_2010}
\bibinfo{author}{\bibfnamefont{J.~F.} \bibnamefont{Sherson}},
  \bibinfo{author}{\bibfnamefont{C.}~\bibnamefont{Weitenberg}},
  \bibinfo{author}{\bibfnamefont{M.}~\bibnamefont{Endres}},
  \bibinfo{author}{\bibfnamefont{M.}~\bibnamefont{Cheneau}},
  \bibinfo{author}{\bibfnamefont{I.}~\bibnamefont{Bloch}}, \bibnamefont{and}
  \bibinfo{author}{\bibfnamefont{S.}~\bibnamefont{Kuhr}},
  \bibinfo{journal}{1006.3799}  (\bibinfo{year}{2010}).

\bibitem[{\citenamefont{Landau and Lifshitz}(1987)}]{landau_fluid_1987}
\bibinfo{author}{\bibfnamefont{L.~D.} \bibnamefont{Landau}} \bibnamefont{and}
  \bibinfo{author}{\bibfnamefont{E.}~\bibnamefont{Lifshitz}},
  \emph{\bibinfo{title}{Fluid Mechanics, Second Edition: Volume 6}}
  (\bibinfo{publisher}{{Butterworth-Heinemann}}, \bibinfo{year}{1987}),
  \bibinfo{edition}{2nd} ed., ISBN \bibinfo{isbn}{0750627670}.

\bibitem[{\citenamefont{Pitaevskii and
  Lifshitz}(1981)}]{pitaevskii_physical_1981}
\bibinfo{author}{\bibfnamefont{L.~P.} \bibnamefont{Pitaevskii}}
  \bibnamefont{and} \bibinfo{author}{\bibfnamefont{E.}~\bibnamefont{Lifshitz}},
  \emph{\bibinfo{title}{Physical Kinetics: Volume 10}}
  (\bibinfo{publisher}{{Butterworth-Heinemann}}, \bibinfo{year}{1981}), ISBN
  \bibinfo{isbn}{0750626356}.

\bibitem[{\citenamefont{Fisher et~al.}(1989)\citenamefont{Fisher, Weichman,
  Grinstein, and Fisher}}]{fisher_boson_1989}
\bibinfo{author}{\bibfnamefont{M.~P.~A.} \bibnamefont{Fisher}},
  \bibinfo{author}{\bibfnamefont{P.~B.} \bibnamefont{Weichman}},
  \bibinfo{author}{\bibfnamefont{G.}~\bibnamefont{Grinstein}},
  \bibnamefont{and} \bibinfo{author}{\bibfnamefont{D.~S.}
  \bibnamefont{Fisher}}, \bibinfo{journal}{Phys. Rev. B}
  \textbf{\bibinfo{volume}{40}}, \bibinfo{pages}{546{\textendash}570}
  (\bibinfo{year}{1989}).

\bibitem[{\citenamefont{Jaksch et~al.}(1998)\citenamefont{Jaksch, Bruder,
  Cirac, Gardiner, and Zoller}}]{jaksch_cold_1998}
\bibinfo{author}{\bibfnamefont{D.}~\bibnamefont{Jaksch}},
  \bibinfo{author}{\bibfnamefont{C.}~\bibnamefont{Bruder}},
  \bibinfo{author}{\bibfnamefont{J.~I.} \bibnamefont{Cirac}},
  \bibinfo{author}{\bibfnamefont{C.~W.} \bibnamefont{Gardiner}},
  \bibnamefont{and} \bibinfo{author}{\bibfnamefont{P.}~\bibnamefont{Zoller}},
  \bibinfo{journal}{Phys. Rev. Lett.} \textbf{\bibinfo{volume}{81}},
  \bibinfo{pages}{3108{\textendash}3111} (\bibinfo{year}{1998}).

\bibitem[{\citenamefont{Zwerger}(2003)}]{zwerger_mott-hubbard_2003}
\bibinfo{author}{\bibfnamefont{W.}~\bibnamefont{Zwerger}},
  \bibinfo{journal}{Journal of Optics B: Quantum and Semiclassical Optics}
  \textbf{\bibinfo{volume}{5}}, \bibinfo{pages}{S9} (\bibinfo{year}{2003}).

\bibitem[{\citenamefont{Press et~al.}(2007)\citenamefont{Press, Teukolsky,
  Vetterling, and Flannery}}]{press_numerical_2007}
\bibinfo{author}{\bibfnamefont{W.~H.} \bibnamefont{Press}},
  \bibinfo{author}{\bibfnamefont{S.~A.} \bibnamefont{Teukolsky}},
  \bibinfo{author}{\bibfnamefont{W.~T.} \bibnamefont{Vetterling}},
  \bibnamefont{and} \bibinfo{author}{\bibfnamefont{B.~P.}
  \bibnamefont{Flannery}}, \emph{\bibinfo{title}{Numerical Recipes 3rd Edition:
  The Art of Scientific Computing}} (\bibinfo{publisher}{Cambridge University
  Press}, \bibinfo{year}{2007}), \bibinfo{edition}{3rd} ed., ISBN
  \bibinfo{isbn}{0521880688}.

\bibitem[{\citenamefont{Wernsdorfer et~al.}(2010)\citenamefont{Wernsdorfer,
  Snoek, and Hofstetter}}]{wernsdorfer_lattice-ramp-induced_2010}
\bibinfo{author}{\bibfnamefont{J.}~\bibnamefont{Wernsdorfer}},
  \bibinfo{author}{\bibfnamefont{M.}~\bibnamefont{Snoek}}, \bibnamefont{and}
  \bibinfo{author}{\bibfnamefont{W.}~\bibnamefont{Hofstetter}},
  \bibinfo{journal}{Phys. Rev. A} \textbf{\bibinfo{volume}{81}},
  \bibinfo{pages}{043620} (\bibinfo{year}{2010}).

\bibitem[{\citenamefont{Menotti and Trivedi}(2008)}]{menotti_spectral_2008-2}
\bibinfo{author}{\bibfnamefont{C.}~\bibnamefont{Menotti}} \bibnamefont{and}
  \bibinfo{author}{\bibfnamefont{N.}~\bibnamefont{Trivedi}},
  \bibinfo{journal}{Phys. Rev. B} \textbf{\bibinfo{volume}{77}},
  \bibinfo{pages}{235120} (\bibinfo{year}{2008}).

\bibitem[{\citenamefont{Landau and Lifshitz}(1981)}]{landau_quantum_1981}
\bibinfo{author}{\bibfnamefont{L.~D.} \bibnamefont{Landau}} \bibnamefont{and}
  \bibinfo{author}{\bibfnamefont{L.~M.} \bibnamefont{Lifshitz}},
  \emph{\bibinfo{title}{Quantum Mechanics {Non-Relativistic} Theory, Third
  Edition: Volume 3}} (\bibinfo{publisher}{{Butterworth-Heinemann}},
  \bibinfo{year}{1981}), \bibinfo{edition}{3rd} ed., ISBN
  \bibinfo{isbn}{0750635398}.

\bibitem[{\citenamefont{Vishveshwara and
  Lannert}(2008)}]{vishveshwara_josephson_2008}
\bibinfo{author}{\bibfnamefont{S.}~\bibnamefont{Vishveshwara}}
  \bibnamefont{and} \bibinfo{author}{\bibfnamefont{C.}~\bibnamefont{Lannert}},
  \bibinfo{journal}{Phys. Rev. A} \textbf{\bibinfo{volume}{78}},
  \bibinfo{pages}{053620} (\bibinfo{year}{2008}).

\end{thebibliography}
\end{document}